\documentclass[a4paper,12pt,preprint,tightenlines,floatfix,showpacs,preprintnumbers,
nofootinbib,eqsecnum]{article}

\usepackage[utf8]{inputenc}
\usepackage{amssymb,amsmath,mathrsfs}
\usepackage[english]{babel}

\usepackage[a4paper, total={7in, 9in}]{geometry}
\usepackage{color}
\usepackage{epstopdf}
\usepackage{graphicx,wrapfig}
\usepackage{multirow}
\usepackage{mathtools}
\usepackage{tabularx}
\usepackage{makecell}
\usepackage{caption}
\usepackage{float}
\usepackage[colorlinks,citecolor=blue,urlcolor=blue,bookmarks=false
]{hyperref}
\RequirePackage{flushend}
\RequirePackage[numbers,sort&compress]{natbib}
\usepackage{verbatim}
\usepackage{authblk}
\usepackage[font=footnotesize,labelfont=bf,singlelinecheck=false]{caption}
\usepackage[binary-units]{siunitx}
\usepackage{tikz}
\usetikzlibrary{arrows,shadows}
\usepackage{setspace}

\newcommand{\msbar}{\overline{\rm{MS}}}

\pdfoutput=1

\begin{document}
\pagenumbering{gobble}
\date{\today}

\title{\vspace{-1.5cm}\begin{flushright}
\small{INR-TH-2022-023}
\end{flushright} \vspace{1.5cm} \textbf{Representation of the RG-invariant quantities in perturbative QCD through powers of the conformal anomaly
}}

\author[1]{ A.~L.~Kataev\footnote{kataev@ms2.inr.ac.ru}}
\author[1, 2, 3, 4]{ V.~S.~Molokoedov\footnote{viktor\_molokoedov@mail.ru}}

\affil[1]{{\footnotesize Institute for Nuclear Research of the Russian Academy of Science,
 117312, Moscow, Russia}}
\affil[2]{{\footnotesize Research Computing Center, Moscow State University, 119991, Moscow, Russia}}
\affil[3]{{\footnotesize  Moscow Center for Fundamental and Applied Mathematics, 119992, Moscow, Russia}}
\affil[4]{{\footnotesize Moscow Institute of Physics and Technology, 141700, Dolgoprudny, Moscow Region, Russia}}

\maketitle

\begin{abstract}

In this work we consider the possibility of representing the perturbative series for renormalization group invariant quantities in QCD in the form of their decomposition in powers of the conformal anomaly $\beta(\alpha_s)/\alpha_s$ in the $\msbar$-scheme. We remind that such expansion is possible for the Adler function of the process of $e^+e^-$ annihilation into hadrons and the coefficient function of the Bjorken polarized sum rule for the deep-inelastic electron-nucleon scattering, which are both related by the Crewther-Broadhurst-Kataev relation. In addition, we study the discussed decomposition for the static quark-antiquark Coulomb-like potential, its relation with the quantity defined by the cusp anomalous dimension and the coefficient function of the Bjorken unpolarized sum rule of neutrino-nucleon scattering. In conclusion we also present the formal results of applying this approach to the non-renormalization invariant ratio between the pole and $\msbar$-scheme running mass of heavy quark in QCD and compare them with those already known in the literature. The arguments in favor of the validity of the considered representation 
in powers of $\beta(\alpha_s)/\alpha_s$ for all mentioned perturbative quantities are discussed.

\end{abstract}


\newpage

\section{Preliminaries}
\pagenumbering{arabic}
\pagestyle{plain}
\setcounter{page}{1}

A long time ago, R.J. Crewther proved in his work \cite{Crewther:1972kn} that in the Born approximation the product of the expressions for the  Adler function of the process $e^+e^- \rightarrow \gamma^* \rightarrow {\textit{hadrons}}$ annihilation in massless limit and the coefficient function of the  Bjorken polarized sum rule of deep-inelastic lepton-nucleon scattering is proportional to amplitude of $\pi^0\rightarrow \gamma\gamma$ decay, which contains the number of quark colors $N_c$. In the normalization of the Adler function by unity, this product also becomes equal to unity. However, when higher orders of the perturbation theory (PT) in powers of the running strong coupling constant $\alpha_s(Q^2)$, defined in the Euclidean region in the MS-like renormalization scheme, are taken into account, the Crewther relation is  modified. Indeed,
starting from $\mathcal{O}(\alpha^2_s)$ level in the product of the flavor non-singlet (NS) perturbative contributions to the Adler function $D_{NS}(Q^2)$ and the coefficient function $C_{NS}(Q^2)$ of the  Bjorken polarized sum rule  instead of unity an extra term appears. It reflects the violation of the symmetry  with respect to conformal transformations of the triangle Green function, composed from axial-vector-vector (AVV) quark currents \cite{Schreier:1971um, Adler:1972msp}. Application of these transformations and the operator product expansion approach to the AVV diagram enabled Crewther to obtain his original result \cite{Crewther:1972kn}.

In the $\msbar$-scheme (and generally speaking in the class of the gauge-invariant MS-like schemes) at $Q^2=\mu^2$ this extra conformal symmetry breaking term is proportional to the renormalization group (RG) $\beta$-function:
\begin{equation}
\label{beta-exp}
\beta^{(M)}(a_s(\mu^2)) = \mu^2 \frac{\partial a_s(\mu^2)}{\partial \mu^2} =
- \sum\limits_{i = 0}^{M-1} \beta_{i} a^{\;i+2}_s(\mu^2). 
\end{equation}

The solution of this RG equation allows to obtain the expression for $a_s(Q^2)=\alpha_s(Q^2)/\pi$ in terms of the renormalized strong coupling 
$a_s(\mu^2)=\alpha_s(\mu^2)/\pi$ depending on the scale $\mu$. The index $M\geq 1$ in Eq.(\ref{beta-exp})  is introduced for convenience and throughout this work it will stand for the order of  approximation in which the concrete quantity will be considered. Note here that one-, two- and three-loop coefficients of $\beta$-function, used by us in this manuscript, were calculated in  \cite{Gross:1973id, Politzer:1973fx}, \cite{Jones:1974mm, Caswell:1974gg, Egorian:1978zx}, \cite{Tarasov:1980au, Larin:1993tp}  correspondingly.

For the first time the modification form of the Crewther relation in the perturbative QCD was discovered at $M=3$ in \cite{Broadhurst:1993ru} and confirmed at $M=4$ later on in \cite{Baikov:2010je}. At least at this level its generalization, called now by the Crewther-Broadhurst-Kataev (CBK) relation, reads
\begin{equation} 
\label{BK}
D^{(M)}_{NS} (a_s)C^{(M)}_{NS}(a_s)=
1+\bigg(\frac{\beta^{(M-1)}(a_s)}{a_s}\bigg)K^{(M-1)}(a_s)+\mathcal{O}(a^{M+1}_s),
\end{equation}
where $a_s=a_s(\mu^2=Q^2)$ and therefore all RG logarithms are nullified. 

Thus, in higher orders of PT the radiative corrections violates the simple Crewther identity and leads to the factorization of the conformal anomaly term $\beta(a_s)/a_s$ in Eq.(\ref{BK}). Vice versa, in the conformal-invariant limit \cite{Schreier:1971um, Adler:1972msp} (and in the scale-invariant one in particular) the $\beta$-function is vanished and the simple Crewther identity is restored. 

The conformal symmetry breaking (CSB) term in the right side of Eq.(\ref{BK}) contains the flavor $n_f$-dependent coefficient function $K^{(M-1)}(a_s)$, which is the $(M-1)$-degree polynomial in $a_s$. The explicit expressions for this function were obtained at $M=3$ in \cite{Broadhurst:1993ru} and at $M=4$ in \cite{Baikov:2010je}.

One should mention that as was recently shown in Ref.\cite{Chetyrkin:2022fqk} the analog of the CBK-relation also holds in the extended QCD model with arbitrary number of fermion representations at $M=4$. This fact demonstrates that the factorization of the $\beta$-function in Eq.({\ref{BK}) is 
not accidental at least at this level.

At the next stage, it was demonstrated in \cite{Kataev:2010dm} at $M=3$ and in \cite{Kataev:2010du} at $M=4$ that in the case of a generic simple gauge group the CSB term can be represented in the expansion form in powers of the conformal anomaly. Then, the CBK relation is rewritten as:
\begin{equation} 
\label{BK-two-fold}
D^{(M)}_{NS} (a_s)C^{(M)}_{NS}(a_s)=
1+\sum\limits_{n=1}^{M-1}\bigg(\frac{\beta^{(M-n)}(a_s)}{a_s}\bigg)^nP^{(M-n)}_n(a_s) +\mathcal{O}(a^{M+1}_s).
\end{equation} 
At the 4-loop level the polynomials $P^{(r)}_n(a_s)$ in Eq.(\ref{BK-two-fold}) read:
\begin{equation}
\label{polynomial}
P^{(r)}_n(a_s)=\sum\limits_{k=1}^r P^{(r)}_{n, k} \;a^{\;k}_s=\sum_{p = 1}^{4-n} a^{\;p}_s \sum_{k=1}^p P_n^{(r)}[k,p-k] C_F^k C_A^{p-k},  
\end{equation}
where $r=4-n$ when $M=4$ and $C_F$, $C_A$ are the quadratic Casimir operator in the fundamental and adjoint representation of the gauge group respectively. Coefficients $P^{(r)}_{n, k}$ are defined unambiguously \cite{Kataev:2010du}. An important point here is that at this level of PT {\textit{all dependence on}} $n_f$ in r.h.s. of Eq.(\ref{BK-two-fold}) is contained in the coefficients of the $\beta$-function. Thus, in contrast to the coefficients of the polynomial $K^{(M-1)}(a_s)$ in Eq.(\ref{BK}), the terms of $P^{(r)}_n(a_s)$ in Eq.(\ref{BK-two-fold}) are independent on the number of quark flavors. 

The double sum representation (\ref{BK-two-fold}), (\ref{polynomial}) for the CSB term motivated the authors of the work 
\cite{Cvetic:2016rot} to consider the similar one for the NS contributions to the Adler function and the coefficient function of the Bjorken polarized sum rule at least at the analytically available $\mathcal{O}(a^4_s)$ level. For instance, according to this paper at $M=4$ the PT expression for the NS Adler function, calculated in the $\msbar$-scheme for the non-abelian gauge theory with a simple compact Lie group, may be presented in the following form:
\begin{equation}
\label{DD-two-fold}
D^{(M)}_{NS}(a_s) = 1 + D^{(M)}_0(a_s)+\sum\limits_{n=1}^{M-1}\left(\frac{\beta^{(M-n)}(a_s)}{a_s}\right)^{n} D^{(M-n)}_{n}(a_s),
\end{equation}
where polynomials $D^{(r)}_n(a_s)$ in the coupling constant $a_s$ are:
\begin{equation}
\label{Drn}
D^{(r)}_n(a_s)=\sum\limits_{k=1}^r D^{(r)}_{n,k}a^{\;k}_s.
\end{equation}

In a more detailed form Eq.(\ref{DD-two-fold}) may be written down as:
\begin{align}
\label{explicit}
\hspace{-0.5cm}
D^{(M=4)}_{NS}(a_s)&=1 + D^{(4)}_{0,1}a_s+\bigg(D^{(4)}_{0,2}-\beta_0 D^{(3)}_{1,1}\bigg)a^2_s+\bigg(D^{(4)}_{0,3}-\beta_0 D^{(3)}_{1,2}-\beta_1 D^{(3)}_{1,1}+\beta^2_0 D^{(2)}_{2,1}\bigg)a^3_s \\ \nonumber
&+\bigg(D^{(4)}_{0,4}-\beta_0 D^{(3)}_{1,3}-\beta_1 D^{(3)}_{1,2}-\beta_2 D^{(3)}_{1,1}+\beta^2_0 D^{(2)}_{2,2}+2\beta_0\beta_1 D^{(2)}_{2,1}-\beta^3_0 D^{(1)}_{3,1}\bigg)a^4_s,
\end{align}
where at the fixed number of $n$ and $k$,  $D^{(r)}_{n,k}\equiv D^{(r+1)}_{n,k}\equiv D^{(r+2)}_{n,k}\equiv \dots$, e.g. the terms $D^{(2)}_{1,2}\equiv D^{(3)}_{1,2}$.

The coefficients $D^{(r)}_{n,k}$ including in Eq.(\ref{explicit}) are determined by an unambiguous way
as solutions of a system of linear equations, analogous to those presented in \cite{Kataev:2010du}. Herewith, the full dependence on $n_f$ (except for the light-by-light scattering effects -- see explanations below) is absorbed into the coefficients of $\beta$-function and their combinations (\ref{explicit}). For the first time at the four-loop level the values of $D^{(r)}_{n,k}$ were obtained in Ref.\cite{Cvetic:2016rot}. It is clear that the representation (\ref{DD-two-fold}) and its counterpart for the coefficient function of the Bjorken polarized sum rule is \textit{a sufficient condition} for the validity of the generalized Crewther relation in the form of Eq.(\ref{BK-two-fold}).

Note here two interesting facts. First of them is that the representation (\ref{DD-two-fold}), whose detailed form at the $\mathcal{O}(a^4_s)$ level is given in Eq.(\ref{explicit}), is in full agreement with its $\{\beta\}$-expansion structure proposed in Ref.\cite{Mikhailov:2004iq} much earlier than this expansion in powers of conformal anomaly. Indeed, accordingly to this work the coefficients $d_M$ $(1\leq M\leq 4)$ of the NS Adler function
\begin{equation}
\label{simD}
D^{(M=4)}_{NS}(a_s)=1+d_1a_s+d_2a^2_s+d_3a^3_s+d_4a^4_s
\end{equation}
may be decomposed into the coefficients of the $\beta$-function and have the following form:
\begin{subequations}
\begin{align}
\label{dd1}
d_1 &= d_1[0], \\ 
\label{dd2}
d_2 &= \beta_0 d_2[1] + d_2[0],  \\ 
\label{dd3}
d_3 &= \beta_0^2 d_3[2] + \beta_1 d_3[0,1] + \beta_0 d_3[1] + d_3[0],  \\ 
\label{dd4}
d_4 &= \beta_0^3 d_4[3] + \beta_1 \beta_0 d_4[1,1] + \beta_2 d_4[0,0,1] + \beta_0^2 d_4[2] + \beta_1  d_4[0,1] + \beta_0 d_4[1] + d_4[0]. 
\end{align}
\end{subequations}
Relations (\ref{dd1}-\ref{dd4}) are in full compliance with representation (\ref{explicit}).

The second interesting fact lies in observation that certain $\beta$-dependent terms $D^{(r)}_{n,k}$ in (\ref{explicit}) {\textit{coincide}} in different orders of PT. For instance, $D^{(3)}_{1,1}$ term is the same for $\beta_0a^2$, $\beta_1a^3_s$ and $\beta_2a^4_s$ contributions. Matching this fact with results of $\{\beta\}$-expansion (\ref{dd2}-\ref{dd4}), one can obtain the following equalities:
\begin{subequations}
\begin{align}
\label{D_{1,1}}
D_{1,1}^{(1)} &= -d_2[1] = -d_3[0,1] = -d_4[0,0,1],  \\
\label{D_{1,2}}
D_{1,2}^{(2)} &= -d_3[1] = -d_4[0,1],  \\
\label{D_{2,1}}
D_{2,1}^{(1)} &= d_3[2] =  d_4[1,1]/2.
\end{align}
\end{subequations}

The $\{\beta\}$-expansion terms (\ref{dd1}-\ref{dd4}) were obtained in  
\cite{Cvetic:2016rot} (see also \cite{Goriachuk:2021ayq}, where the misprints made in \cite{Cvetic:2016rot} were found and several $\{\beta\}$-expanded coefficients of $d_5$ were fixed as well). As a result, the application of Eqs.(\ref{D_{1,1}}-\ref{D_{2,1}}) enables to determine all terms of $\{\beta\}$-expansion, except for ones proportional to $\beta^0_0$ and higher powers of $\beta_0$ in concrete order of PT. Indeed, all other terms  proportional to higher coefficients of $\beta$-function will be uniquely fixed from $\{\beta\}$-expansion pattern obtained at the previous order of PT. In the case, when the coefficient $d_M$ is known explicitly, the terms proportional to powers of $\beta_0$ are defined unambiguously.


It is worth pointing out that the same relations between some $\{\beta\}$-dependent terms were also observed and used e.g. in Refs.\cite{Mojaza:2012mf, Brodsky:2013vpa, Wu:2019mky} in the process of application of the $\mathcal{R}_\delta$-scheme (or class of the MS-like schemes) to the study of the $\{\beta\}$-pattern of the perturbative series for different observables in QCD. As was explained there, the application of the $\mathcal{R}_\delta$-scheme exposes
a special degeneracy of the $\{\beta\}$-expanded terms in various orders of PT. Now we see that this degeneracy may be clearly understood from the decomposition of the PT series for physical quantities in powers of $\beta(\alpha_s)/\alpha_s$.

Let us raise the following issue: is this decomposition valid for arbitrary observables that are not related by the CBK relation? In order to shed light on this question, we will consider the  higher order PT corrections to various physical quantities and apply to them the $\{\beta\}$-expansion technique in powers of $\beta(\alpha_s)/\alpha_s$ in next sections. Further, we will compare some of our results with those already known in literature, obtained by independent methods.

\section{The case of the static potential}

\subsection{The expansion in powers of the conformal anomaly}

Let us start our consideration from the case of the static interaction potential of the heavy quark and antiquark in QCD in color-singlet state. The perturbative Coulomb-like part of this quantity is defined through the vacuum expectation value of the gauge-invariant Wilson loop $W[C]$:
\begin{align}
\label{V-def}
&V_{Q\bar{Q}}(r)=-\lim\limits_{T\rightarrow \infty}\frac{1}{iT}\log W[C]= \int \frac{d^3\vec{q}}{(2\pi)^3}e^{i\vec{q}\vec{r}}V_{Q\bar{Q}}(\vec{q}^{\;2}),
\\
\label{W}
&\text{where} ~~ W[C]=\frac{\langle 0|~{\rm{Tr ~\hat{P}}}\exp\bigg(ig\oint\limits_{C} dx^\mu A^a_\mu t^a\bigg)~|0\rangle}{\langle 0| ~{\rm{Tr}} ~1 ~|0\rangle}
\end{align}
and $\vec{q}^{\;2}$ is the square of the Euclidean three-dimensional momentum,  $C$ is a closed rectangular contour, $T$ and $r$ are the time and three-dimensional spatial variables, $\hat{P}$ is the path-ordering operator, $A^a_\mu$ is a gluon field and $t^a$ are the generators of the Lie algebra of the generic simple gauge group in fundamental representation. In our study we are primarily interested in the case of the $SU(N_c)$ gauge color group. 

The limit $T\rightarrow\infty$ formally leads to $q_0\rightarrow 0$ and the square of the Euclidean four-dimensional transferred momentum $Q^2\rightarrow \vec{q}^{\;2}$. Thus, technically we carry out the transition from the Euclidean four-dimensional space to its three-dimensional subspace. The $\msbar$-scheme $\beta$-function is also considered in this subspace and the coupling $a_s(\vec{q}^{\;2})$ starts to depend on $\vec{q}^{\;2}$.

The perturbative expression for the Fourier image $V_{Q\bar{Q}}(\vec{q}^{\;2})$  is known in analytic form in the $\rm{\overline{MS}}$ renormalization scheme 
at the three-loop 
level (see \cite{Fischler:1977yf, Billoire:1979ih}, 
\cite{Peter:1996ig, Schroder:1998vy} and \cite{Smirnov:2008pn, Lee:2016cgz} respectively). It is written as:
\begin{equation}
\label{V}
V_{Q\bar{Q}}(\vec{q}^{\; 2})=-\frac{4\pi C_F\alpha_s(\vec{q}^{\; 2})}{\vec{q}^{\; 2}}\bigg[1+a_1a_s(\vec{q}^{\; 2})+a_2a^2_s(\vec{q}^{\; 2})  
+ \bigg(a_3+\frac{\pi^2 C^3_A L}{8}\bigg)a^3_s(\vec{q}^{\; 2})+\mathcal{O}(a^4_s)\bigg].
\end{equation}
Here $L=\log(\mu^2/\vec{q}^{\; 2})$ and the term $\pi^2C^3_AL/8$ \cite{Brambilla:1999qa} arises due to the infrared divergences. 

Let us introduce the next notation:
\begin{equation}
\label{mathscr-V}
\mathscr{V}_{Q\bar{Q}}(a_s(\vec{q}^{\; 2}))= 1+a_1a_s(\vec{q}^{\; 2})+a_2a^2_s(\vec{q}^{\; 2})  
+a_3a^3_s(\vec{q}^{\; 2}) + \mathcal{O}(a^4_s).
\end{equation}

Following our idea of decomposition of the perturbative coefficients into powers of the conformal anomaly, which is valid for the ones of the Adler function and the coefficient function of the Bjorken polarized sum rule at least at $M=4$ in QCD, we can write down the expression for $\mathscr{V}_{Q\bar{Q}}(a_s(\vec{q}^{\; 2}=\mu^2))$ in the $M$-th order of PT ($1\leq M\leq 3$) in the following form:
\begin{eqnarray}
\label{V-beta}
&&\mathscr{V}^{(M)}_{Q\bar{Q}}(a_s)=1+\mathscr{V}^{(M)}_0(a_s)+\sum\limits_{n=1}^M\bigg(\frac{\beta^{(M-n+1)}(a_s)}{a_s}\bigg)^n\mathscr{V}^{(M-n+1)}_n(a_s), \\
&&\text{where} ~~~~~\mathscr{V}^{(M)}_0(a_s)=\sum\limits_{k=1}^{M}\mathscr{V}^{(M)}_{0,k}a^k_s, ~~~~~~~ \mathscr{V}^{(r)}_n(a_s)=\sum\limits_{k=1}^{r}\mathscr{V}^{(r)}_{n,k}a^{k-1}_s.
\end{eqnarray}

In a more detailed way, Eq.(\ref{V-beta}) may be rewritten as:
\begin{align}
\label{V-det-f}
\hspace{-0.5cm}
\mathscr{V}^{(M=3)}_{Q\bar{Q}}(a_s)&=1 +\bigg(\mathscr{V}^{(3)}_{0,1}-\beta_0 \mathscr{V}^{(3)}_{1,1}\bigg)a_s+\bigg(\mathscr{V}^{(3)}_{0,2}-\beta_0 \mathscr{V}^{(3)}_{1,2}-\beta_1 \mathscr{V}^{(3)}_{1,1}+\beta^2_0 \mathscr{V}^{(2)}_{2,1}\bigg)a^2_s \\ \nonumber
&+\bigg(\mathscr{V}^{(3)}_{0,3}-\beta_0 \mathscr{V}^{(3)}_{1,3}-\beta_1 \mathscr{V}^{(3)}_{1,2}-\beta_2 \mathscr{V}^{(3)}_{1,1}+\beta^2_0 \mathscr{V}^{(2)}_{2,2}+2\beta_0\beta_1 \mathscr{V}^{(2)}_{2,1}-\beta^3_0 \mathscr{V}^{(1)}_{3,1}\bigg)a^3_s,
\end{align}
where at the fixed numbers $n$ and $k$, $\mathscr{V}^{(r)}_{n,k}\equiv\mathscr{V}^{(r+1)}_{n,k}\equiv\mathscr{V}^{(r+2)}_{n,k}\equiv\dots$. As we have anticipated, the representation (\ref{V-det-f}) is in full agreement with its $\{\beta\}$-expansion pattern
\begin{subequations}
\begin{align}
\label{a_1-beta}
a_1&=\beta_0a_1[1]+a_1[0], \\
\label{a_2-beta}
a_2&=\beta^2_0a_2[2]+\beta_1a_2[0,1]+\beta_0a_2[1]+a_2[0], \\
\label{a_3-beta}
a_3&=\beta^3_0a_3[3]+\beta_1\beta_0a_3[1,1]+\beta_2a_3[0,0,1]+\beta^2_0a_3[2]+\beta_1a_3[0,1]+\beta_0a_3[1]+a_3[0].
\end{align}
\end{subequations}

Moreover, the representation (\ref{V-det-f}) provides us equalities similar to Eqs.(\ref{D_{1,1}}-\ref{D_{2,1}}), namely:
\begin{subequations}
\begin{align}
\label{1}
a_1[1]&= a_2[0,1]=a_3[0,0,1]=-\mathscr{V}^{(1)}_{1,1}, \\
\label{2}
a_2[1]&= a_3[0,1]=-\mathscr{V}^{(2)}_{1,2}, \\
\label{3}
a_2[2]&= a_3[1,1]/2=\mathscr{V}^{(2)}_{2,1}.
\end{align}
\end{subequations}

Using the three-loop explicit analytic results \cite{Lee:2016cgz} for the static potential in the perturbative QCD and taking into account Eqs.(\ref{1}-\ref{3}), we obtain all terms of $\{\beta\}$-expansion included in Eqs.(\ref{a_1-beta}-\ref{a_3-beta}). The corresponding results are  given in Table \ref{T-1}.

Note that there are transcendent numbers $\alpha_4={\rm{Li}}_4(1/2)+\log^4 2/4!$ with polylogarithmic function ${\rm{Li}}_{n}(x)=\sum_{k=1}^{\infty}x^kk^{-n}$, $s_6=\zeta_6+\zeta_{-5,-1}$ with $\zeta_6=\pi^6/945$ and multiple zeta value $\zeta_{-5,-1}=\sum_{k=1}^{\infty}\sum_{i=1}^{k-1}(-1)^{i+k}/ik^5$ in Table \ref{T-1}. They appear from scalar master-integrals \cite{Lee:2016cgz}, which are different from those arising in the calculation of corrections to $D_{NS}(Q^2)$ and $C_{NS}(Q^2)$ (see e.g. \cite{Baikov:2010hf}).

Excepting two reservations, the presented coefficients of $\{\beta\}$-expansion mostly coincide with ones defined in \cite{Brodsky:2013vpa} by independent way using the $\mathcal{R}_\delta$-scheme approach. The first of them lies in the small difference between our rational numbers, highlighted in blue in Table \ref{T-1}, and those found in \cite{Brodsky:2013vpa}. Indeed, instead of coefficients $5171/288$ and $-2981/192 \approx -15.526$, contained in $a_3[2]$-term in Ref.\cite{Brodsky:2013vpa}, we get slightly different ones, namely $5471/288$ and $-7943/576 \approx -13.789$. Instead of the $C_FC_A$-term $-66769/3456\approx -19.320$ included in $a_3[1]$  in \cite{Brodsky:2013vpa}, we obtain $-70069/3456\approx -20.275$.

The second clause concerns the light-by-light scattering effects, first occurring at the three-loop level for the static potential \cite{Lee:2016cgz}. In our opinion, despite the $n_f$-dependence of these effects, they have to be included in the scale-invariant coefficient $a_3[0]$. This fact is reflected in Table \ref{T-1} by the 
incorporation of the terms proportional to $d^{abcd}_Fd^{abcd}_Fn_f/N_A$ and $d^{abcd}_Fd^{abcd}_A/N_A$ group structures in this ``$n_f$-independent'' coefficient. Here $N_A$ is the number of group generators, $d^{abcd}_F={\rm{Tr}}(t^at^{\mathop{\{}b}t^ct^{d\mathop{\}}})/6$ and $d^{abcd}_A={\rm{Tr}}(C^aC^{\mathop{\{}b}C^cC^{d\mathop{\}}})/6$, where
the symbol $\{\dots\}$ stands for the full symmetrization procedure of elements $t^bt^ct^d$ by superscripts $b$, $c$ and $d$ \cite{vanRitbergen:1997va, Czakon:2004bu};  $(C^a)_{bc}=-if^{abc}$ are the generators of the adjoint representation with the antisymmetric structure constants $f^{abc}$ of the Lie algebra: $[t^a,t^b]=if^{abc}t^c$. In the particular case of the $SU(N_c)$  color gauge group the contractions of the symmetric invariant tensors considered above read:
\begin{gather}
\label{color-structures-1}
\frac{d^{abcd}_Fd^{abcd}_F}{N_A}=\frac{N^4_c-6N^2_c+18}{96N^2_c}, ~~~~~ \frac{d^{abcd}_Fd^{abcd}_A}{N_A}=\frac{N_c(N^2_c+6)}{48}.
\end{gather}

\begin{table}[H]
\renewcommand{\tabcolsep}{0.6cm} 
\renewcommand{\arraystretch}{1.7}
\centering
\vspace{-1cm}
\begin{tabular}{|c|c|c|}
\hline
Coefficients              & Group structures &                                                  Numbers   
\\ \hline
$a_1[1]$   & \textemdash  & $\frac{5}{3}$                          
\\ \hline
 $a_1[0]$  &    $C_A$     & $-\frac{2}{3}$ 
\\ \hline
$a_2[2]$   & \textemdash  & $\frac{25}{9}$                            
\\ \hline                                        
\multirow{2}{*}{$a_2[1]$} & $C_F$ & $\frac{35}{16}-3\zeta_3$    \\ \cline{2-3} 
                          & $C_A$         & $-\frac{217}{72}+\frac{7}{2}\zeta_3$ 
\\ \hline
\multirow{2}{*}{$a_2[0]$} & $C_FC_A$ & $-\frac{385}{192}+\frac{11}{4}\zeta_3$    \\ \cline{2-3} 
                          & $C^2_A$         & $\frac{133}{144}-\frac{11}{4}\zeta_3+\frac{\pi^2}{4}-\frac{\pi^4}{64}$ 
\\ \hline
$a_3[3]$   & \textemdash  & $\frac{125}{27}$                            
\\ \hline
\multirow{2}{*}{$a_3[2]$} & $C_F$ & ${\color{blue}{\frac{5471}{288}}}-\frac{39}{2}\zeta_3$    \\ \cline{2-3} 
                          & $C_A$         & ${\color{blue}{-\frac{7943}{576}}}+\frac{69}{4}\zeta_3+\frac{\pi^4}{15}$ 
\\ \hline
& $C^2_F$          & $-\frac{571}{192}-\frac{19}{8}\zeta_3+\frac{15}{2}\zeta_5$                                     \\ \cline{2-3} 
& $C_FC_A$       & ${\color{blue}{-\frac{70069}{3456}}}+\frac{49}{2}\zeta_3-\frac{15}{4}\zeta_5$               \\ \cline{2-3} 
\multirow{-3}{*}{$a_3[1]$}
& $C^2_A$   & \begin{tabular}[c]{@{}c@{}} $\frac{2491}{288}-\frac{309}{16}\zeta_3-\frac{1091}{128}\zeta_5-\frac{171}{128}\zeta^2_3+\frac{9}{4}s_6-\frac{761}{53760}\pi^6$ \\
$+\pi^4\bigg(\frac{9}{640}+\frac{5}{192}\log 2-\frac{3}{64}\log^2 2\bigg)$ \\
$+\pi^2\bigg(-\frac{17}{576}+\frac{19}{64}\zeta_3+\frac{1}{16}\log 2+\frac{21}{32}\zeta_3\log 2+\frac{3}{2}\alpha_4\bigg)$
\end{tabular}              
\\ \hline
& $C^2_FC_A$          & $\frac{6281}{2304}+\frac{209}{96}\zeta_3-\frac{55}{8}\zeta_5$                                     \\ \cline{2-3} 
& $C_FC^2_A$          & $\frac{3709}{3456}-\frac{379}{96}\zeta_3+\frac{55}{16}\zeta_5$                                     \\ \cline{2-3} 
\multirow{4}{*}{$a_3[0]$}
& $C^3_A$   & \begin{tabular}[c]{@{}c@{}} $-\frac{19103}{27648}+\frac{181}{48}\zeta_3+\frac{1431}{512}\zeta_5+\frac{55}{512}\zeta^2_3+\frac{3}{16}s_6-\frac{21097}{1935360}\pi^6$ \\
$+\pi^4\bigg(\frac{211}{23040}-\frac{15}{256}\log 2-\frac{61}{2304}\log^2 2\bigg)$ \\
$+\pi^2\bigg(-\frac{191}{768}+\frac{841}{768}\zeta_3-\frac{955}{576}\log 2+\frac{203}{384}\zeta_3\log 2+\frac{5}{3}\alpha_4\bigg)$
\end{tabular}  
\\ \cline{2-3}
& $\frac{d^{abcd}_Fd^{abcd}_F}{N_A}n_f$ & \begin{tabular}[c]{@{}c@{}} $\frac{5}{96}\pi^6+\pi^4\bigg(-\frac{23}{24}+\frac{1}{6}\log 2-\frac{1}{2}\log^2 2\bigg)$ \\
$+\pi^2\bigg(\frac{79}{36}-\frac{61}{12}\zeta_3+\log 2+\frac{21}{2}\zeta_3\log 2\bigg)$
\end{tabular}    
\\ \cline{2-3}
& $\frac{d^{abcd}_Fd^{abcd}_A}{N_A}$ & \begin{tabular}[c]{@{}c@{}} $\frac{1511}{2880}\pi^6+\pi^4\bigg(-\frac{39}{16}+\frac{35}{12}\log 2+\frac{31}{12}\log^2 2\bigg)$ \\
$+\pi^2\bigg(\frac{929}{72}-\frac{827}{24}\zeta_3-74\alpha_4+\frac{461}{6}\log 2-\frac{217}{4}\zeta_3\log 2\bigg)$
\end{tabular}
\\ \hline
\end{tabular}
\captionsetup{justification=centering}
\caption{\label{T-1} All terms included in the $\{\beta\}$-expansion of coefficients $a_1$, $a_2$ and $a_3$.}
\end{table}

Although the term $d_F^{abcd}d_F^{abcd}n_f/N_A$ in the correction $a_3$ (\ref{mathscr-V})  is proportional
to the number of flavors $n_f$, which formally enters the $\beta_0$-coefficient, we will not include it
into $\mathscr{V}^{(3)}_{1,3}$-coefficient in Eq.(\ref{V-det-f}), since such rearrangement will not be supported by the QED limit \cite{Cvetic:2016rot, Goriachuk:2021ayq}. Indeed, in the QED limit of the QCD-like theory with the $SU(N_c)$ group $d_F^{abcd}d^{abcd}_F/N_A=1$ and $n_f=N$, where $N$ is the number of the charged leptons (structures with $d^{abcd}_A$ are nullified). This term arises from the three-loop Feynman diagram with light-by-light scattering internal subgraphs (see Fig.\ref{LBL}). However, in QED the sum of these subgraphs are convergent and does
not give extra $\beta_0$-dependent (or $N$-dependent) contribution to the coefficient $\mathscr{V}^{(3)}_{1,3}$ \cite{Cvetic:2016rot, Goriachuk:2021ayq}. Therefore, to get a smooth transition from the case of $SU(N_c)$ to $U(1)$  gauge group,  these light-by-light scattering terms should be included into the $\beta$-independent coefficient $\mathscr{V}^{(3)}_{0,3}$ (or $a_3[0]$) at the three-loop level.

\begin{figure}[h!]
\centering
\includegraphics[width=0.4\textwidth]{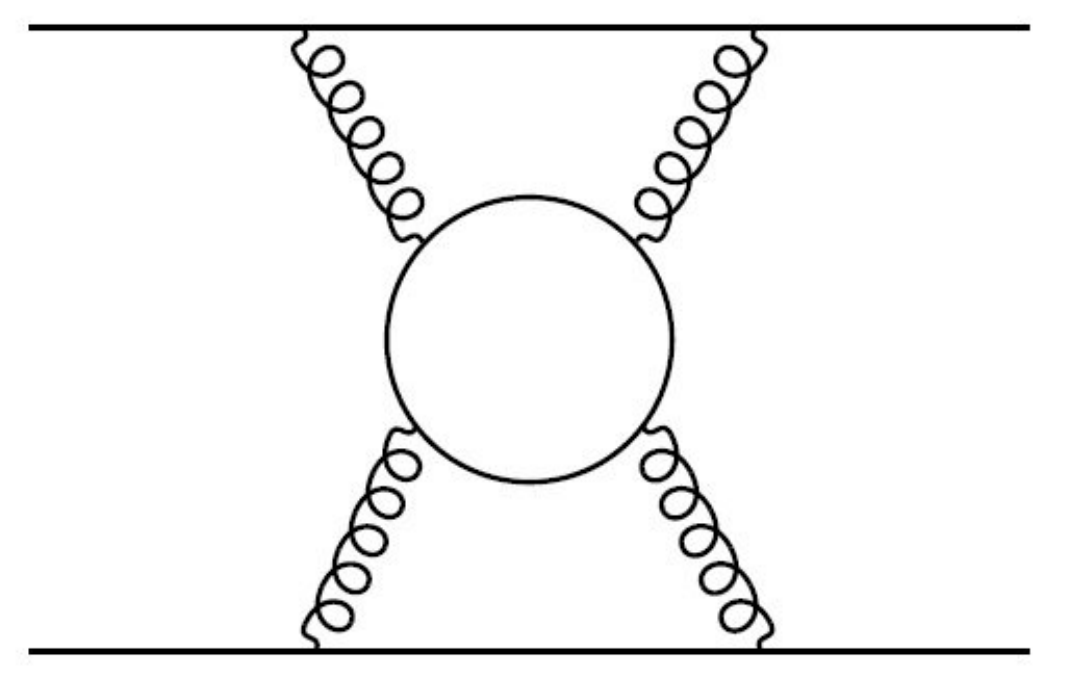}
\centering
\captionsetup{justification=centering}
\caption{Light-by-light scattering type diagram in the static potential}
\label{LBL}
\end{figure}



In accordance with the foregoing, at the three-loop level the static Coulomb potential of QED can be expressed through the invariant charge as:
\begin{equation}
V_{{\rm{QED}}}(\vec{q}^{\; 2})=-\frac{4\pi}{\vec{q}^{\; 2}}\frac{\alpha(\mu^2)}{1+\Pi_{{\rm{QED}}}(L, \alpha(\mu^2))}\bigg(1+N\cdot\mathcal{C}\bigg(\frac{\alpha(\mu^2)}{\pi}\bigg)^3\bigg)+\mathcal{O}(\alpha^5),
\end{equation}
where the $\msbar$-scheme photon vacuum polarization function $\Pi_{{\rm{QED}}}(L, \alpha(\mu^2))$ was calculated in $\alpha^3$ order in Ref.\cite{Gorishnii:1991hw}. The constant $\mathcal{C}$ originates from the light-by-light scattering effects and may be extracted from the results of work \cite{Lee:2016cgz}:
\begin{equation}
\mathcal{C}=\frac{5}{96}\pi^6-\pi^4\bigg(\frac{23}{24}-\frac{1}{6}\log 2+\frac{1}{2}\log^2 2\bigg)+\pi^2\bigg(\frac{79}{36}-\frac{61}{12}\zeta_3+\log 2+\frac{21}{2}\zeta_3\log 2\bigg).
\end{equation}

\subsection{Special representations of the relation between $\mathscr{V}_{Q\bar{Q}}$ and $\mathscr{V}_{{\rm{cusp}}}$}

It is interesting to note that there is a quantity $\mathscr{V}_{{\rm{cusp}}}$  associated with the static potential by the relation where, as in the CBK relation, the effect of conformal symmetry breaking is explicitly manifested in the form of the factorized $\beta$-function. This relation was first obtained at two-loop level (at $M=2$) in Ref.\cite{Grozin:2015kna} and reads:
\begin{equation}
\label{Vcusp-VQQ}
\mathscr{V}^{(M)}_{{\rm{cusp}}}(a_s)-\mathscr{V}^{(M)}_{Q\bar{Q}}(a_s)=\frac{\beta^{(M-1)}(a_s)}{a_s}C^{(M-1)}(a_s).
\end{equation}
Here $C^{(M-1)}(a_s)=\sum\limits_{k=1}^{M-1}C_ka^k_s$ with $C_1=\frac{47}{27}C_A-\frac{28}{27}T_Fn_f$. 

In Eq.(\ref{Vcusp-VQQ}) the function $\mathscr{V}_{{\rm{cusp}}}(a_s)$ is related with the cusp anomalous dimension $\Gamma(\phi, a_s)$ of the vacuum expectation value of the Wilson loop $W[\tilde{C}]$ (\ref{W}), but unlike the case of the static potential, $\tilde{C}$ is now a closed contour formed by two segments along directions $v^\mu_1$ and $v^\mu_2$ with Euclidean cusp angle $\phi$: $\cos \phi=(v_1, v_2)$ (see investigations on this topic \cite{Grozin:2015kna, Grozin:2018vdn, Bruser:2019auj, Henn:2019swt, vonManteuffel:2020vjv, Grozin:2022wse} and references therein).

The quantity $\log W[\tilde{C}]$ is expressed through the logarithm of the corresponding $\msbar$-scheme cusp renormalization constant $Z_{{\rm{cusp}}}$ \cite{Grozin:2015kna, Grozin:2018vdn, Bruser:2019auj, Henn:2019swt, vonManteuffel:2020vjv, Grozin:2022wse}:
\begin{equation}
\log W= \log Z_{{\rm{cusp}}} +\mathcal{O}(\varepsilon^0),
\end{equation}
where $Z_{{\rm{cusp}}}$ contains the ultraviolet divergences and it is written in terms of the renormalized coupling constant. In its turn, the cusp anomalous dimension is defined as:
\begin{equation}
\label{def-Gamma}
\Gamma(\phi, a_s)=\frac{\partial \log Z_{{\rm{cusp}}}}{\partial \log \mu^2}.
\end{equation}

The quantity $\mathscr{V}_{{\rm{cusp}}}(a_s)$ is a $\delta$-independent coefficient function of $\Gamma(\phi, a_s)$ in the anti-parallel lines limit, i.e. when $\phi=\pi-\delta$ and $\delta\ll 1$ \cite{Kilian:1993nk, Grozin:2015kna}:
\begin{equation}
\label{Vcusp-def}
\Gamma(\pi-\delta, \alpha_s)=-C_F\alpha_s\frac{\mathscr{V}_{{\rm{cusp}}}(\alpha_s)}{2\delta}+\mathcal{O}\bigg(\frac{\alpha^4_s\log \delta}{\delta}\bigg).
\end{equation}
The first order corrections to $\mathscr{V}_{{\rm{cusp}}}(\alpha_s)$ and $\mathscr{V}_{Q\bar{Q}}(\alpha_s)$ coincide \cite{Kilian:1993nk} and the differences between them are observed only from the second order of PT  and they are proportional to the first coefficient of the RG $\beta$-function \cite{Grozin:2015kna} (see Eq.(\ref{Vcusp-VQQ})).

At $M=3$ the convincing arguments in favor of  validity of the expression (\ref{Vcusp-VQQ}) were given in \cite{Bruser:2019auj}. Except for the pure non-abelian part proportional to $C^2_A$-structure, all other terms to coefficient $C_2$ (proportional to terms $T^2_Fn^2_f$, $C_FT_Fn_f$ and $C_AT_Fn_f$) were defined there in analytical form.

Since the function $\mathscr{V}_{{\rm{cusp}}}(\alpha_s)$ has the same dependence on $n_f$ and the same group structures like the quantity $\mathscr{V}_{Q\bar{Q}}(\alpha_s)$, it is obvious that its series of PT can be decomposed into powers of the  conformal anomaly as well:
\begin{align}
\label{Vcusp-beta}
\mathscr{V}^{(M)}_{{\rm{cusp}}}(a_s)=1+\mathscr{V}^{(M)}_{{\rm{cusp}}, \;0}(a_s)+\sum\limits_{n=1}^M\bigg(\frac{\beta^{(M-n+1)}(a_s)}{a_s}\bigg)^n\mathscr{V}^{(M-n+1)}_{{\rm{cusp}}, \;n}(a_s).
\end{align}

But representation of the functions $\mathscr{V}_{Q\bar{Q}}(\alpha_s)$  and $\mathscr{V}_{{\rm{cusp}}}(\alpha_s)$ in the form of double sum (\ref{V-beta}) and (\ref{Vcusp-beta}) is a sufficient condition for the CSB term in the difference $\mathscr{V}_{{\rm{cusp}}}(a_s)-\mathscr{V}_{Q\bar{Q}}(a_s)$ can also be decomposed in powers of $\beta(\alpha_s)/\alpha_s$:
\begin{align}
\label{CSBforcuspinbeta}
\mathscr{V}^{(M)}_{{\rm{cusp}}}(a_s)-\mathscr{V}^{(M)}_{Q\bar{Q}}(a_s)&=\sum\limits_{n=1}^M\bigg(\frac{\beta^{(M-n+1)}(a_s)}{a_s}\bigg)^nT^{(M-n+1)}_{n}(a_s), \\
&\text{where} ~~ T^{(r)}_n=\sum\limits_{k=1}^{r}T^{(r)}_{n,k}a^{k-1}_s.
\end{align}

One should emphasize that since the first order corrections to $\mathscr{V}_{{\rm{cusp}}}(a_s)$ and to $\mathscr{V}_{Q\bar{Q}}(a_s)$ coincide, then $T^{(1)}_{1,1}=T^{(2)}_{1,1}=\dots=0$. Taking this fact into account and using the representation (\ref{CSBforcuspinbeta}) and the results of Refs.\cite{Grozin:2015kna, Bruser:2019auj}, we obtain:
\begin{align}
\label{CSBcusp-num}
\mathscr{V}^{(3)}_{{\rm{cusp}}}(a_s)-\mathscr{V}^{(3)}_{Q\bar{Q}}(a_s)&=\bigg(-\frac{28}{9}\beta^2_0+\frac{10}{9}C_A\beta_0\bigg)a^2_s+\bigg(\beta^3_0\bigg(2\zeta_3-\frac{134}{27}\bigg)+\frac{10}{9}C_A\beta_1 \\ \nonumber
&-2\cdot\frac{28}{9}\beta_0\beta_1 
+\beta^2_0\bigg(C_F\bigg(\frac{19}{2}\zeta_3-\frac{1487}{96}+\frac{\pi^4}{20}\bigg)+C_A\bigg(\frac{1879}{144}-\frac{89}{12}\zeta_3-\frac{\pi^4}{8}\bigg)\bigg) \\ \nonumber
&+\beta_0\bigg(C_FC_A\bigg(\frac{16357}{1152}-\frac{209}{24}\zeta_3-\frac{11}{240}\pi^4\bigg)-C^2_AT^{AA}_{1,3} 
\bigg)\bigg)a^3_s,
\end{align}
where $T^{AA}_{1,3}$ is the still unknown coefficient, contained in term $T^{(3)}_{1,3}$ and proportional to $C^2_A$ group factor. Its explicit form may be found from the coefficient $C_2$ \cite{Bruser:2019auj}, where $C^2_A$-term $C^{AA}_2$
was fixed in part. They are related to each other in the following way: $T^{AA}_{1,3}=C^{AA}_2+86893/15552-737/144\zeta_3-11/96\pi^4$. It is important to emphasize that the system of linear equations, arising in process of  application of the decomposition into powers of the conformal anomaly, 
has a unique solution at any value of $T^{AA}_{1,3}$. This means that the representation (\ref{CSBforcuspinbeta}) is always possible at least at the three-loop level.

Note that the absence of terms $\beta_1a^2_s$ and $\beta_2a^3_s$ in Eq.(\ref{CSBcusp-num}) is explained by fact that $T^{(2)}_{1,1}=T^{(3)}_{1,1}=0$.

\section{The case of the unpolarized Bjorken sum rule}

Let us now consider a quantity that, according to current data, is not included in any of known relations that would reflect the effect of violation of the conformal symmetry. As an example, we examine the unpolarized Bjorken sum rule for deep-inelastic neutrino-nucleon scattering \cite{Bjorken:1967px}. The detailed theoretical study of this process may be still of interest in view of its possible investigation from the potential future DIS $\nu N$ data, which may be collected at SAND detector (see e.g. \cite{Vicenzi:2022mqk}) of the DUNE Collaboration.

So, consider the aforementioned unpolarized Bjorken sum rule:
\begin{align}
F_{Bjunp}(Q^2)=\int\limits_0^1 dx (F^{\bar{\nu}p}_1(x, Q^2)-F^{\nu p}_1(x, Q^2)),
\end{align}
where $F^{\bar{\nu}p \;(\nu p)}_1(x, Q^2)$ is the structure function of $\nu N$ DIS process that arises in the general decomposition of the hadronic tensor into possible Lorentz structures. At large $Q^2=-q^2$ the dependence of $F_{Bjunp}$ on $Q^2$ is absorbed into the coupling constant. 

In the Born approximation $F_{Bjunp}=1$. Deviation from unity is already observed in the leading order of PT \cite{Bardeen:1978yd}. Analytical two- and three-loop corrections to the coefficient function of the unpolarized Bjorken sum rule in the $\msbar$-scheme for the case of the generic simple gauge group were computed in \cite{Gorishnii:1983gs} and 
\cite{Larin:1990zw} correspondingly. In particular case of the $SU(3)$ color group the analytical four-loop correction to $F_{Bjunp}$ is known due to the unpublished results \cite{Chetyrkin:2015}. One should mention that the general formula for the leading renormalon contributions to this considered perturbative quantity was obtained in \cite{Broadhurst:2002bi}. Its predictions are in full agreement with the results of the papers cited above.

Applying the idea about the representation of the perturbative series for observables in QCD-like theories with the generic simple gauge group in form of their decomposition into powers of $\beta(\alpha_s)/\alpha_s$, we obtain that at $1 \leq M\leq 4$ the NS contribution to the Bjorken unpolarized sum rule may be rewritten as:
\begin{align}
F^{(M)}_{Bjunp}(a_s)&=1+F^{(M)}_0(a_s)+\sum\limits_{n=1}^{M-1}\bigg(\frac{\beta^{(M-n)}(a_s)}{a_s}\bigg)^n F^{(M-n)}_n(a_s). 
\end{align}

In more detailed form at the four-loop level we have:
\begin{align}
\label{detF}
F^{(M=4)}_{Bjunp}(a_s)&=1 + F^{(4)}_{0,1}a_s+\bigg(F^{(4)}_{0,2}-\beta_0 F^{(3)}_{1,1}\bigg)a^2_s+\bigg(F^{(4)}_{0,3}-\beta_0 F^{(3)}_{1,2}-\beta_1 F^{(3)}_{1,1}+\beta^2_0 F^{(2)}_{2,1}\bigg)a^3_s \\ \nonumber
&+\bigg(F^{(4)}_{0,4}-\beta_0 F^{(3)}_{1,3}-\beta_1 F^{(3)}_{1,2}-\beta_2 F^{(3)}_{1,1}+\beta^2_0 F^{(2)}_{2,2}+2\beta_0\beta_1 F^{(2)}_{2,1}-\beta^3_0 F^{(1)}_{3,1}\bigg)a^4_s 
\end{align}

Taking into account the results of Refs.\cite{Bardeen:1978yd, Gorishnii:1983gs, Larin:1990zw}, one can get:
\begin{subequations}
\begin{align}
F_{0,1}&=-\frac{C_F}{2}, ~~~ F_{1,1}=\frac{4}{3}C_F, ~~~ F_{0,2}=\frac{11}{16}C^2_F-\frac{1}{24}C_FC_A,  ~~~ F_{2,1}=-\frac{155}{36}C_F, \\ 
F_{1,2}&=\bigg(\frac{431}{96}-\frac{1}{2}\zeta_3\bigg)C^2_F+\bigg(-\frac{35}{144}+\frac{7}{2}\zeta_3-5\zeta_5\bigg)C_FC_A, \\ 
F_{0,3}&=\bigg(-\frac{313}{64}-\frac{47}{4}\zeta_3+\frac{35}{2}\zeta_5\bigg)C^3_F+\bigg(\frac{687}{128}+\frac{125}{8}\zeta_3-\frac{95}{4}\zeta_5\bigg)C^2_FC_A \\ \nonumber
&+\bigg(-\frac{463}{288}-\frac{137}{24}\zeta_3+\frac{115}{2}\zeta_5\bigg)C_FC^2_A. 
\end{align}

Using now the results of four-loop calculations \cite{Chetyrkin:2015} made for particular case of the SU(3) color gauge group, we arrive to the following expressions:
\begin{align}
\label{F31}
F_{3,1}&=\frac{1780}{81}, ~~~ F_{2,2}=\frac{78155}{2592}+\frac{87}{2}\zeta_3+12\zeta^2_3-110\zeta_5, \\ 
\label{F13}
F_{1,3}&=-\frac{97247}{2592}-\frac{61153}{648}\zeta_3-\frac{965}{54}\zeta_5+\frac{296}{27}\zeta^2_3-\frac{49}{4}\zeta_7-6B, \\ 
\label{F04}
F_{0,4}&=-\frac{8139161}{124416}-\frac{308489}{1296}\zeta_3+\frac{239665}{1296}\zeta_5-\frac{2927}{108}\zeta^2_3+\frac{253757}{2592}\zeta_7-\frac{33}{2}B+B\cdot n_f,
\end{align}
\end{subequations}
where $B\sim d^{abcd}_Fd^{abcd}_F/d_R$-term is the light-by-light scattering type contribution to the Bjorken unpolarized sum rule. Note that the nonabelian flavor-independent correction of this effect is already contained in the coefficient $F_{0,4}$. To get expressions (\ref{F31}-\ref{F04}) in an analytical form for the case of the generic simple gauge group, it would be interesting to evaluate the 4-loop correction to the coefficient function of the Bjorken unpolarized sum rule for the generic simple gauge group as well.

\section{Conclusion}
The proposed decomposition procedure of the PT series for observables into powers of the conformal anomaly in QCD allows us not only to reproduce the known structure of $\{\beta\}$-expansion, but to predict its definite terms in higher orders of PT as well. For instance, application of this technique to the non-singlet contribution of the coefficient function of the Bjorken unpolarized sum rule gives
\begin{align*}
F^{(M=4)}_{Bjunp}(a_s)&=1 + F^{(4)}_{0,1}a_s+\bigg(F^{(4)}_{0,2}-\beta_0 F^{(3)}_{1,1}\bigg)a^2_s+\bigg(F^{(4)}_{0,3}-\beta_0 F^{(3)}_{1,2}-\beta_1 F^{(3)}_{1,1}+\beta^2_0 F^{(2)}_{2,1}\bigg)a^3_s \\ \nonumber
&+\bigg(F^{(4)}_{0,4}-\beta_0 F^{(3)}_{1,3}-\beta_1 F^{(3)}_{1,2}-\beta_2 F^{(3)}_{1,1}+\beta^2_0 F^{(2)}_{2,2}+2\beta_0\beta_1 F^{(2)}_{2,1}-\beta^3_0 F^{(1)}_{3,1}\bigg)a^4_s.
\end{align*}

We can see that the knowledge of $F^{(3)}_{1,1}$-coefficient of $\beta_0$-dependent term in $a^2_s$ order enables to predict values of $\beta_1$- and $\beta_2$-dependent ones in $a^3_s$ and $a^4_s$ orders. In its turn, the knowledge of $\beta_0$-dependent coefficient $F^{(3)}_{1,2}$ and $\beta^2_0$-dependent one $F^{(2)}_{2,1}$ in $a^3_s$ order gives values of the $\beta_1$- and $\beta_0\beta_1$-dependent coefficients in $a^4_s$ order correspondingly. Thus, the utilization of the expansion in powers of the conformal anomaly allows 
to determine all terms of $\{\beta\}$-expansion, except for ones proportional to $\beta^0_0$ and higher powers of $\beta_0$ in concrete order of PT. Indeed, all other terms  proportional to higher coefficients of $\beta$-function will be uniquely fixed from $\{\beta\}$-expansion pattern obtained at the previous order of PT. In the case, when the correction of the $M$-th order to observable quantity is known explicitly, the terms proportional to powers of $\beta_0$ may be defined unambiguously.

Application of this procedure to the static Coulomb-like potential, to the relation between it and cusp anomalous dimension in anti-parallel lines limit and to the Bjorken unpolarized sum rule enables to determine all their $\{\beta\}$-expanded terms unambiguously. The arguments in favor of validity of this decomposition in powers of $\beta(a_s)/a_s$ gives not only from Refs.\cite{Mojaza:2012mf, Brodsky:2013vpa, Wu:2019mky} and \cite{Kataev:2016aib}, \cite{Goriachuk:2021ayq}, but also from work \cite{Melnikov:2000qh}, where the ratio of the pole mass $M$ to $\msbar$-scheme scale-dependent running $\overline{m}(\mu^2)$ mass of the heavy quark was obtained at the three-loop level analytically.  The results of this work were also presented there in the following numerical form
\begin{equation}
\label{M-ratio}
\frac{M}{\overline{m}(\overline{m}^2)}=1+1.333a_s+(6.248\beta_0-3.739)a^2_s+(23.497\beta^2_0+6.248\beta_1+1.019\beta_0-29.94)a^3_s.
\end{equation}

It reproduces the $\{\beta\}$-expansion structure and the relations between its definite coefficients, which we have obtained within procedure of decomposition in powers of $\beta(a_s)/a_s$ (see e.g. Eqs.(\ref{D_{1,1}}-\ref{D_{2,1}}) or (\ref{1}-\ref{3})). Indeed, the coefficient before $\beta_0a^2_s$-term is the same as the coefficient before $\beta_1a^3_s$-term. The same feature we have observed above in the cases of the NS  Adler function, the NS coefficient functions of the Bjorken polarized and unpolarized sum rules and the static Coulomb-like potential. However, unlike the previous cases, the scale dependence of ratio $M/\overline{m}(\mu^2)$ is governed now by two RG functions: $\beta$-function and anomalous mass dimension $\gamma_m$. Note one interesting fact. The utilization of \textit{the formal}
$\{\beta\}$-\textit{expansion in powers of} $\beta(a_s)/a_s$ of the two- \cite{Tarrach:1980up, Nachtmann:1981zg} and three-loop \cite{Tarasov:1982gk, Larinmass} coefficients of $\gamma_m$-function, namely
\begin{subequations}
\begin{align}
\label{gamma1an}
\gamma_1&=\frac{5}{8}C_F\beta_0+\frac{3}{32}C^2_F+\frac{7}{16}C_FC_A, \\
\label{gamma2an}
\gamma_2&=-\frac{35}{48}C_F\beta^2_0+\beta_0\bigg(C^2_F\bigg(\frac{27}{16}-\frac{9}{4}\zeta_3\bigg)+C_FC_A\bigg(-\frac{679}{96}+\frac{9}{4}\zeta_3\bigg)\bigg)+\frac{5}{8}C_F\beta_1+\frac{129}{128}C^3_F \\ \nonumber
&+C^2_FC_A\bigg(-\frac{525}{256}+\frac{33}{16}\zeta_3\bigg)+C_FC^2_A\bigg(\frac{1063}{128}-\frac{33}{16}\zeta_3\bigg),
\end{align}
\end{subequations}
and the subsequent substitution of these decompositions into NLO and NNLO general expressions, given in Ref.\cite{Huang:2022rij} and obtained there within the $\mathcal{R}_\delta$-scheme motivated approach, allows us to reproduce the numerical form of Eq.(\ref{M-ratio}). This finding may be related to the fact that the QCD trace anomaly $T^{\mu}_{\mu}=(\beta(a_s)/2a_s)F^{a\mu\nu}F^a_{\mu\nu}+(1+\gamma_m)m\bar{\psi}\psi$  \cite{Adler:1976zt, Collins:1976yq, Nielsen:1977sy} contains not only $\beta$-term, but also the anomalous mass dimension $\gamma_m$-term.
However, as was noted in \cite{Kataev:2016aib},
it is necessary to treat with caution to the $\{\beta\}$-expansion of the non-RG-invariant quantities with other non-zero anomalous dimension functions.

\section*{Acknowledgments}

We would like to thank S.V. Mikhailov for useful discussions and comments. The work of VSM was supported by the Russian Science Foundation, agreement no. 21-71-30003 (study of the possibility of representing the perturbative series for RG invariant quantities in powers of the conformal anomaly) and by the Ministry of Education and Science of the Russian Federation as part of the program of the Moscow Center for Fundamental and Applied Mathematics, agreement No. 075-15-2019-1621 (analytical analysis for the case of the generic simple gauge group).

\begin{flushleft}

\end{flushleft}

\end{document}